\documentclass[journal]{IEEEtran}
\usepackage{amsmath}%
\usepackage{amsfonts}%
\usepackage{amssymb}%
\usepackage{graphicx}
\usepackage{url}
\usepackage{booktabs}
\usepackage{multirow}
\usepackage{placeins}
\usepackage{comment}
\usepackage{todonotes}
\usepackage{url}
\usepackage{pifont}
\usepackage{subcaption}

\usepackage{bm}
\usepackage{textcomp}

\usepackage{xcolor}
\usepackage{enumitem}

\usepackage{math/math_defs}

\usepackage{array}
\newcolumntype{P}[1]{>{\centering\arraybackslash}p{#1}}

\usepackage{caption}

\begin{document}
\title{Adversarial Attacks and Defenses\\ for Speech Recognition Systems}

\author{Piotr Żelasko,~\IEEEmembership{Member,~IEEE,}
        Sonal Joshi,~\IEEEmembership{Student Member,~IEEE,}
        Yiwen Shao,~\IEEEmembership{Student Member,~IEEE,}
        Jesús Villalba,~\IEEEmembership{Member,~IEEE,}
        Jan Trmal,~\IEEEmembership{Senior~Member,~IEEE,}/
        Najim Dehak,~\IEEEmembership{Senior Member,~IEEE,}
        and~Sanjeev Khudanpur,~\IEEEmembership{Senior Member,~IEEE}%
        
\thanks{All authors were with the Center of Language and Speech Processing, The John's Hopkins University, Baltimore,
MD, 21210 USA. E-mail: piotr.andrzej.zelasko@gmail.com.}
\thanks{
This project was supported by DARPA Award HR001119S0026-GARD-FP-052}
\thanks{Manuscript received March 4, 2021.}}

\markboth{Submission under review}
{Żelasko \MakeLowercase{\textit{et al.}}: Adversarial Attacks and Defenses for Speech Recognition Systems}

\maketitle

\begin{abstract}
The ubiquitous presence of machine learning systems in our lives necessitates research into their vulnerabilities and appropriate countermeasures.
In particular, we investigate the effectiveness of adversarial attacks and defenses against automatic speech recognition (ASR) systems.
We select two ASR models -- a thoroughly studied DeepSpeech model and a more recent Espresso framework Transformer encoder-decoder model.
We investigate two threat models: a denial-of-service scenario where fast gradient-sign method (FGSM) or weak projected gradient descent (PGD) attacks are used to degrade the model's word error rate (WER); and a targeted scenario where a more potent imperceptible attack forces the system to recognize a specific phrase.
We find that the attack transferability across the investigated ASR systems is limited.
To defend the model, we use two preprocessing defenses: randomized smoothing and WaveGAN-based vocoder, and find that they significantly improve the model's adversarial robustness. 
We show that a WaveGAN vocoder can be a useful countermeasure to adversarial attacks on ASR systems -- even when it is jointly attacked with the ASR, the target phrases' word error rate is high.
\end{abstract}


%
\IEEEpeerreviewmaketitle

\section{Introduction}

\IEEEPARstart{A}{dversarial} attacks aim to fool a machine learning system by perturbing the inputs in such a way to affect its decision while not altering that of a human~\cite{szegedy-iclr14}. 
With an ever-increasing presence of automation in our lives, vulnerabilities to such attacks pose a significant threat to our society. 
There is an extensive body of research on how to attack artificial intelligence systems, e.g.~\cite{Goodfellow2015,carlini-16,Kurakin2017,Dong2019}.
In this work, we focus mainly on adversarial attacks against automatic speech recognition (ASR) systems and investigate how to defend them.

Most studies attack end-to-end ASR systems like Mozilla's DeepSpeech~\cite{amodei2016deep}.
For example, the study~\cite{Cisse2017} attacks DeepSpeech using the Houdini attack, an attack tailored for task performance measure that can be applied to multiple areas, including speech recognition. Another study~\cite{carlini-18} proposes a white-box iterative optimization-based adversarial attack (henceforth referred to as Carlini-Wagner attack), demonstrating a 100\% success rate.  \cite{Neekhara2019} propose an audio-agnostic universal adversarial perturbation for DeepSpeech. The authors in~\cite{iter-17} attack WaveNet~\cite{oord2016wavenet} using fast gradient sign method~\cite{goodfellow_generative_2014} and the fooling gradient sign method~\cite{szegedy-iclr14}. Other works attack the state-of-the-art ASR based on Kaldi~\cite{povey2011kaldi}. The authors in~\cite{yuan-18} propose a surreptitious attack to a Kaldi-based ASR by embedding voice into songs and playing it in the background while being inaudible to the human ear.~\cite{Schonherr2019, Qin2019} show that psychoacoustic modeling can be leveraged to make the attacks imperceptible.
While all these are simulated software attacks, \cite{Qin2019} and \cite{Yakura2019} show that physical adversarial attacks, meaning adversarial attacks generated over-the-air using realistic simulated environmental distortions, also break the ASR system. A more detailed overview of adversarial attacks and countermeasures on ASR is presented in~\cite{wang2020adversarial}.

Defending a machine learning system against adversarial attacks is inherently difficult given the wide variety of attack methods.
Two common strategies that were previously proposed for ASR are preprocessing defenses and adversarial training. 
Preprocessing defenses transform the input signal (waveform or features) before passing to the ASR system, hoping that the adversarial noise is lost in the process. 
Examples of preprocesing defenses are audio turbulence and audio squeezing~\cite{yuan-18}; local smoothing using filter, down-sampling and recovery, and quantization~\cite{chen2019real}; MP3 compression~\cite{rigoll2020mp3}.
\cite{joshi2021adversarial} proposed manifold-based defenses for audio -- with WaveGAN vocoder amongst them -- that cast the attacked input signal onto a latent manifold of clean examples, and then reconstruct it. 

Adversarial training entails training the models with data augmented using adversarial perturbations. They can be estimated using projected gradient descent (PGD)~\cite{pal2020adversarial}, as well as using multi-task objectives, feature-scattering (FS), and margin losses~\cite{pal2020adversarial}.
However, the caveat of adversarial training is that it can be susceptible to attack algorithms and threat models unseen during training~\cite{zhang2019limitations,laidlaw2020perceptual}, and careful training hyperparameter tuning is required. 
Given these issues, our work focuses on the more generalized class of preprocessing defenses, where the adversarial attacks are tackled before the classifier.
The main advantage of these defenses is that we do not need information about the type of attacks an adversary can use.

We study two ASR systems with significantly different architectures -- DeepSpeech and Espresso Transformer, described in Section~\ref{sec:asr}.
We explore two threat models - a denial-of-service attack that is supposed to degrade the ASR performance and a targeted attack that forces the ASR to recognize the phrase dictated by the attacker, regardless of what was spoken.
The attack methods are presented in Section~\ref{sec:attacks}. 
We investigate the efficacy of three defenses: randomized smoothing, a WaveGAN vocoder~\cite{joshi2021adversarial}, and label smoothing described in Section~\ref{sec:defenses}. 
The experimental setup and results are shown in Sections~\ref{sec:exp}~and~\ref{sec:results}.
We conclude this work in Section~\ref{sec:conclusions}

\section{Automatic speech recognition systems}
\label{sec:asr}

\textbf{DeepSpeech 2}~\cite{amodei2016deep} consists of several convolutional and recurrent layers and is trained with the CTC criterion~\cite{graves2006connectionist}. The model is frame-synchronous, meaning that each speech frame is assigned a label. DeepSpeech performs character-level recognition\footnote{\url{https://github.com/hkakitani/deepspeech.pytorch}}. 
We do not use an external language model for inference-time decoding.
DeepSpeech has been extensively used for ASR adversarial attack research in the past~\cite{Cisse2017,carlini-18,Neekhara2019}, but it has fallen behind the ASR state-of-the-art in recent years. Therefore, we introduce a second, more recent model for comparison.

\textbf{Espresso} is a framework for training encoder-decoder models for ASR~\cite{wang2019espresso} that is based on fairseq~\cite{ott2019fairseq}. Encoder-decoder models are frame-asynchronous. The encoder produces a sequential latent representation of the input that the decoder autoregressively attends to in multiple decoding iterations, typically resulting in a much shorter output sequence. 
We use the Transformer architecture~\cite{vaswani2017attention}, which has recently been shown to outperform RNN-based models in the speech recognition task~\cite{karita2019comparative}. Our experiments are based on the Espresso \textit{asr\_librispeech} example, with adjustments for on-the-fly differentiable feature extraction that allows to back-propagate to the time-domain audio signal\footnote{\url{https://github.com/pzelasko/espresso/tree/feature/librispeech-wav-model}}.

\section{Adversarial attacks}
\label{sec:attacks}

We consider two threat models. 
In the first one, the attacker creates an additive noise of low magnitude to mislead the ASR system into recognizing a completely different phrase than the one uttered. 
This approach could be described as denial-of-service (DOS). 
In the second, the noise is crafted more carefully to make the ASR recognize a specific target phrase.

We present three methods of estimating additive noise. 
These methods are white-box attacks, meaning that the attacker needs to have access to a copy of the model's weights.
All of these methods are considered targeted attacks -- in order to estimate them, we provide an alternative transcript as the target for the objective function computation when back-propagating.
Note that even if a target phrase is provided for some attacks, the attack may not succeed in making the system recognize it but could still severely modify the ASR output, effectively turning a targeted attack into a DOS attack.

\textbf{Fast gradient sign method (FGSM)}~\cite{Goodfellow2015} takes the benign audio waveform of length $T$ $\xvec\in\real^T$  and computes an adversarial example $\xvec^\prime$ by taking a single step in the direction that minimizes the loss w.r.t. the attacker's target phrase:
\begin{align}
    \label{eq:fgsm}
    \xvec^\prime = \xvec + \varepsilon\,\sign(\nabla_\xvec L(g(\xvec),y^{\mathrm{target}})) \;,
\end{align}
where function $g(\xvec)$ is the ASR model producing a sequence of symbols $y^{\mathrm{predicted}}$, $L$ is categorical cross-entropy loss, $y^{\mathrm{target}}$ is the attacker's target phrase. $\varepsilon$ restricts the  $L_\infty$ norm of the perturbation by imposing $ \lVert \xvec^\prime - \xvec \rVert_{\infty}
\leq \varepsilon $ to keep the attack imperceptible. In other words, the larger the epsilon, the more perceptible are the perturbations and the more effective are the attacks (meaning the ground truth WER deteriorates and the target WER possibly decreases).

\textbf{Projected gradient descent}~\cite{madry2018towards} takes iterative smaller steps $\alpha$ in the direction of the gradient in contrast to FGSM which takes a single step
\begin{align}
    \label{eq:iterfgsm}
   \xvec^{\prime}_{i+1} = \xvec + \clip_\varepsilon(
   \xvec^{\prime}_{i} + \alpha\,\sign(\nabla_{\xvec^{\prime}_{i}} L(g(\xvec^{\prime}_{i}), y^{\mathrm{target}}))) \;,
\end{align}
where $\xvec^{\prime}_0=\xvec$ and $i$ is iteration step for optimization. The $\clip$ function (projection) assures that the $L_\infty$ norm of perturbation is smaller than $\varepsilon$ after each optimization step $i$. This results in a stronger attack than that of FGSM, however it also takes more time to compute. 

\textbf{Imperceptible attack}, introduced by~\cite{Qin2019}, optimizes for two objectives. 
The first objective is to estimate an attack that fools the network (with the same as in equation~\ref{eq:iterfgsm}).
The second objective uses a psychoacoustic model to make the adversarial perturbation imperceptible due to frequency masking~\cite{lin2015principles}. The detailed equations solving the optimization process are presented in~\cite{Qin2019}.

\section{Defense methods}
\label{sec:defenses}

\textbf{Randomized smoothing} is a preprocessing defense that tries to mask the adversarial signal with additive random, normally-distributed noise. 
The noise's standard deviation $\sigma$ is a hyperparameter that controls the trade-off between robustness and accuracy. 
Randomized smoothing is a certifiable defense against attacks with bounded $L_2$ norm perturbations. 
In~\cite{cohen2019certified}, the authors prove tight bounds for certified accuracies under Gaussian noise smoothing. Despite this defense not being certifiable for $L_p$ with $p\ne2$, we found that it also performs well for other norms like $L_\infty$.

The main requirement for this defense to be effective is that the model needs to be robust to Gaussian noise.
The baseline ASR systems are hardly noise-robust, given that LibriSpeech -- with mostly good quality recordings -- is used as the training set with no data augmentation. 
Hence, we make the ASR more robust by introducing randomized smoothing as a data augmentation technique.
 
\textbf{WaveGAN vocoder}~\cite{joshi2021adversarial} reconstructs the speech waveform given a compressed representation of it -- in our case, the log-Mel-spectrograms. We used the ParallelWaveGAN architecture originally proposed in~\cite{yamamoto2020parallel} that acts as a generator in a generative adversarial network (GAN). This vocoder is trained on a combination of waveform-domain adversarial loss; and a reconstruction loss in Short Time Fourier Transform (STFT) domain. The reconstruction loss improves the stability and efficiency of the adversarial training. It is the sum of several STFT losses computed with different spectral analysis parameters (window length/shift, FFT length). The generator's architecture is a non-autoregressive WaveNet, while the architecture for the discriminator is based on a dilated convolutional network.

We considered two alternatives regarding the knowledge of the attacker. \emph{Black-box WaveGAN} where the attacker does not have access to the WaveGAN model; 
and \emph{White-box WaveGAN} where the attacker can backpropagate through the WaveGAN model to jointly attack the ASR and the defense.

\textbf{Label smoothing} is a commonly used technique in training time to improve the general performance of neural networks~\cite{szegedy2016rethinking}. It uses smoothed uniform label vectors in place of one-hot vectors in the cross-entropy computation. In recent works, label smoothing has also been reported to be an effective defense method against adversarial examples on images~\cite{warde201611}. 
Since it is also a part of standard Espresso recipes, all our results except one are reported with label smoothing included. To check its effect on the system's robustness, we run a single ablation experiment by re-training the model without label smoothing.

\section{Experimental setup}
\label{sec:exp}

We train both systems on full LibriSpeech~\cite{panayotov2015librispeech} 960 hours corpus, with no data augmentation (except for the randomized smoothing, which is used as a defense). The systems are evaluated on the first 100 utterances from the \textit{test-clean} split due to the computational complexity of attacking and defending the systems. Unless explicitly stated, both DeepSpeech and Espresso Transformer are trained using their standard LibriSpeech recipes.

For FGSM and PGD attacks, we check their effectiveness at various max-norm levels: 0.0001, 0.001, 0.01, 0.1, and 0.2. For PGD, we use 7 iterations and a learning rate five times lower than the max-norm. These attacks use random target phrases sampled from LibriSpeech training set close to the ground truth transcript's length.

For the two-stage imperceptible target attack \cite{Qin2019}, we have different settings on DeepSpeech and Espresso systems to get low target WERs as baselines. For DeepSpeech, we set the initial max-norm in the first stage to be 0.01 and then gradually reduce it during optimization following \cite{carlini-18} with a decay factor of 0.5. The max iterations and learning rates for the two stages are (100, 8e-4) and (25, 8e-7), respectively. For Espresso, the initial max-norm is set to 0.75, and the learning rates in the two stages are 0.001 and 1e-7. Similar to \cite{Qin2019}, we did all our target attack experiments on 100 pairs of utterances with a matched length of transcription. 

For randomized smoothing defense, we use Gaussian noise with standard deviation $\sigma=0.01$ during test time. We also evaluated $\sigma=0.001$ and noticed little effect on the systems, and $\sigma=0.1$ that completely broke both systems (WER $\geq 70\%$). When used as a data augmentation technique, we add a Gaussian noise with randomly selected $\sigma$ in the range of 0-0.3 to the training recordings. For the ParallelWaveGAN vocoder, we trained a model on random 100k utterances from LibriSpeech training set\footnote{\url{https://github.com/kan-bayashi/ParallelWaveGAN}}.

To measure how successful are the attacks and defenses, we need to check the WER under three conditions:
\begin{enumerate}
    \item \textbf{Clean WER}, measured w.r.t. the true transcript when a system is not attacked but may be defended. It is a measure of how the defense affects the system in its normal operating conditions.
    \item \textbf{Ground-truth WER} measured w.r.t. the true transcript when a system is under attack. It measures the degradation of - a possibly defended - system's performance when under attack.
    \item \textbf{Target WER} measured w.r.t the attacker's targeted transcript when a system is under attack. It shows whether the attackers are able to achieve their goals.
\end{enumerate}

We report the target WER only for the Imperceptible attack since it is always close to 100\% in our FGSM and PGD experiments. We expect it is possible to obtain a lower value with a greater number of PGD iterations, although that would have been redundant given a well-tuned Imperceptible attack. Therefore, we interpret the FGSM and PGD attacks in this work as denial-of-service threats and the Imperceptible attack as an actual targeted attack.

\section{Results}
\label{sec:results}

\begin{table*}
\centering
\caption{\label{tab:wer_results}
Word error rate (\%) for DeepSpeech 2 and Espresso Transformer ASR systems under various attack and defense combinations. We provide both the ground-truth WER (GT) and target WER (TGT) for the Imperceptible attack. \textit{RS$\sigma$} stands for randomized smoothing with a $\sigma$ parameter, \textit{RSAUG} means randomized smoothing augmentation in training, \textit{WAVEAUG} means that the training data was first re-synthesized using WaveGAN.
}
\begin{tabular}{@{}lccccccccccccc@{}}
\toprule
\textbf{Architecture}  & \textbf{Clean}  & \multicolumn{5}{c}{\textbf{FGSM Attack}} & \multicolumn{5}{c}{\textbf{PGD Attack}} & \multicolumn{2}{c}{\textbf{Imperceptible}} \\
\cmidrule(r){1-1}\cmidrule(lr){2-2}\cmidrule(lr){3-7}\cmidrule(lr){8-12}\cmidrule(l){13-14}
$L_{\infty}$ (max-norm) &  & 0.0001 & 0.001 & 0.01 & 0.1 & 0.2  & 0.0001 & 0.001 & 0.01 & 0.1 & 0.2 & GT & TGT \\
\midrule
(1) DeepSpeech 2 & 10.9 & 14.5 & 34.6 & 77.9 & 100.0 & 99.8 & 15.2 & 46.4 & 97.5 & 111.5 & 111.1 & 102.4 & 5.1 \\
(2) \quad + \textit{RS0.01} & 20.9 & 21.3 & 27.6 & 80.6 & 105.9 & 103.4 & 21.2 & 31.3 & 90.6 & 114.6 & 112.9  & 98.8 & 80.1\\
(3) \quad + \textit{RSAUG} & 13.2 & 21.7 & 28.1 & 69.5 & 114.7 & 116.3 & 22.4 & 30.1 & 87.5 & 118.7 & 122.5 & 97.2	& 82.6 \\
(4) \quad + \textit{RSAUG, RS0.01} & 14.7 & 24.2 & 29.2 & 69.1 & 113.4 & 112.0 & 25.4 & 32.3 & 85.6 & 116.6 & 119.7 & 97.0 & 86.7 \\
(5) \quad + \textit{WaveGAN} & 14.1 & 15.2 & 17.1 & 73.8 & 101.3 & 100.2 & 15.9 & 16.0 & 38.4 & 105.4 & 104.8 & 48.0 & 101.8 \\
(6) \quad + \textit{RSAUG, WaveGAN} & 17.0 & 27.2 & 28.3 & 34.0 & 110.9 & 115.1 & 30.4 & 31.4 & 32.1 & 99.7 & 122.9   & 37.2 & 102.4 \\
(7) \quad + \textit{RSAUG, WaveGAN, RS0.01} & 20.2 & 31.6 & 30.4 & 36.0 & 108.6 & 113.5 & 32.9 & 32.4 & 34.2 & 98.3 & 122.3 & 39.2 & 101.6 \\
\midrule
\textit{Ablation: (6) + WAVEAUG} & 11.3 & 13.5 & 14.1 & 65.2 & 99.0 & 99.7 & 13.1 & 12.9 & 31.4 & 102.9 & 102.9 & 42.3 & 102.0 \\
\textit{Ablation: (6) + whitebox WaveGAN} & 17.5 & 30.5 & 32.9 & 47.5 & 125.3 & 118.9 & 32.9 & 44.2 & 76.7 & 124.5 & 133.5 & 97.6 & 93.0 \\
\midrule
(8) Espresso Transformer & 4.4 & 4.7 & 10.2 & 65.3 & 108.0 & 105.0 & 4.7 & 7.9 & 22.6 & 97.3 & 139.3 & 102.6 & 4.0 \\
(9) \hskip1.5em\relax + \textit{RS0.01} & 11.9 & 13.2 & 12.7 & 40.3 & 102.4 & 191.4 & 12.9 & 11.7 & 33.3 & 105.6 & 102.6 & 103.0 & 24.7 \\
(10) \quad + \textit{RSAUG} & 5.6 & 5.5 & 6.0 & 25.4 & 122.8 & 121.1 & 5.5 & 6.1 & 18.1 & 81.2 & 108.2 & 99.7 & 12.6 \\
(11) \quad + \textit{RSAUG, RS0.01} & 5.6 & 5.7 & 6.1 & 16.4 & 94.2 & 110.5 & 5.9 & 6.6 & 15.9 & 124.1 & 124.1 & 98.7 & 26.4 \\
(12) \quad + \textit{WaveGAN} & 5.3 & 5.7 & 5.9 & 11.2 & 97.4 & 119.5 & 5.3 & 6.1 & 19.6 & 99.8 & 104.1 & 37.4 & 100.1 \\
(13) \quad + \textit{RSAUG, WaveGAN} & 5.2 & 6.0 & 6.0 & 7.7 & 70.4 & 101.5 & 5.9 & 6.2 & 7.4 & 83.5 & 114.1 & 15.8 & 101.7 \\
(14) \quad + \textit{RSAUG, WaveGAN, RS0.01} & 6.6 & 6.2 & 6.7 & 10.5 & 96.1 & 116.1 & 6.3 & 6.6 & 7.1 & 78.2 & 108.1 & 18.3 & 102.3 \\
\midrule
\textit{Ablation: (8) without label smoothing} & 4.1 & 4.4 & 8.3 & 22.2 & 99.3 & 485.7 & 4.9 & 11.5 & 58.6 & 108.2 & 131.0 & 102.9 & 5.8 \\
\midrule
\textit{Transfer attack: (1) + attack from (8)} & 10.9 & - & 15.1 & - & - & - & - & 13.7 & - & - & - & 81.2 & 99.5 \\
\textit{Transfer attack: (8) + attack from (1)} & 4.4 & - & 5.41 & - & - & - & - & 8.0 & - & - & - & 17.9 & 102.3 \\
\bottomrule
\end{tabular}
\end{table*}

We present the results of our evaluations in Table~\ref{tab:wer_results}. 

\textbf{Undefended systems.} We find that Espresso (8) achieves more than twice lower WER than DeepSpeech (1) in clean (un-attacked) conditions. Under FGSM attack, DeepSpeech is affected even by 1e-4 max-norm noise (+50\,\% relative WER) and breaks completely at 1e-2 (78\% absolute WER) and above. Espresso exhibits more robustness to small perturbations - 1e-4 max-norm noise increases the WER by 7\% relative, but with greater max-norms, the trend is the same as with DeepSpeech. Under the PGD attack, the difference between the two systems is much more pronounced. The weakest PGD attack completely breaks DeepSpeech, but Espresso appears more robust to PGD than to FGSM. Since label smoothing is a part of standard Espresso training recipe, we retrained Espresso without it to see if the robustness can be attributed to that. Indeed, without label smoothing, we observe that Espresso becomes more vulnerable to PGD than to FGSM. Finally, both systems can be successfully attacked with the Imperceptible attack, yielding a target WER of 4-5\,\%.

\textbf{Randomized smoothing} hurts the baseline systems performance (rows 2 and 9). However, its application in training improves the system's robustness both to random and adversarial noise (rows 3, 4, 10, and 11) at the cost of 15\,\% and 27\,\% relative word error rate increase. Interestingly, Espresso degraded more in the clean condition, but unlike DeepSpeech, it was able to further benefit from inference-time randomized smoothing (rows 11 vs. 10). Randomized smoothing to some extent disturbed the targeted attack on DeepSpeech (target WER increased from 5.1\% to 80-86\%), but not as much on Espresso (4.0\% to 12-26\%), while the ground truth WER remained at 100\%.

\textbf{WaveGAN as a preprocessing defense} significantly improves the adversarial robustness of both DeepSpeech and Espresso (rows 5 and 12) for all types of attacks. Unlike smoothing, it can thwart the targeted Imperceptible attack (target WER increases to 100\% in both systems) but does not fully recover the ground truth transcript (ground-truth WER is reduced from 100\% to 48\% in DeepSpeech and 37.4\% in Espresso). However, it also degrades the un-attacked system performance - DeepSpeech is affected more than Espresso. 

\textbf{Combining inference-time WaveGAN with training-time random smoothing augmentation}, we observe different behavior in the two systems (rows 6 and 13). DeepSpeech's performance is significantly worse than for pure WaveGAN. We suspect that DeepSpeech cannot generalize well to the type of distortions introduced by WaveGAN. To that end, we performed an ablation study where we trained DeepSpeech on LibriSpeech recordings re-synthesized by the WaveGAN and random smoothing augmentation. It indeed improved the performance in most conditions, confirming our hypothesis. On the other hand, Espresso did not require additional tuning to leverage WaveGAN and achieved WER improvements in most attack conditions, including a ground-truth 15.8\% WER in the Imperceptible attack. We see that this combination of defenses starts to help even for high max-norm attacks, such as FGSM with 0.1 max-norm. We find that adding random smoothing on top of WaveGAN during inference does not help the systems further - both seem to suffer both in un-attacked and attacked conditions (rows 7 and 14).

\textbf{White-box WaveGAN} stands for vocoder jointly attacked with the ASR system. Since DeepSpeech is much smaller than Espresso, we study that system alone due to the experiment's computational complexity. We find that FGSM and PGD with a small number of iterations degrade the ASR performance more (especially 0.01 max-norm PGD). The Imperceptible attack is partially defended - the reference transcript is not recognized, but neither is the attacker's target phrase.

\textbf{Attack transferability.} We find limited evidence of attack transferability for the investigated systems. FGSM and PGD performance are much worse than for the systems they were estimated on (except for already low Espresso PGD WER). The imperceptible attack fails to recognize the target phrases, but severely hurts DeepSpeech performance, which can also be explained by the attack's high magnitude when estimated on Espresso.

\section{Conclusion}
\label{sec:conclusions}

We presented our evaluation of the adversarial robustness of two vastly different ASR systems, DeepSpeech 2 and Espresso.
We confirmed that both systems are vulnerable to every adversarial attack in the study when no counter-measures are applied to protect them, but Espresso exhibits significantly more inherent robustness to lower max-norm perturbations.
We provide evidence that attack transferability between these models is limited.
We showed that high max-norm perturbations (0.1 and greater) devastate the ASR performance and found no defense helpful in these circumstances.
We found randomized smoothing mostly helpful as a data augmentation technique rather than a preprocessing defense, and even then, its effectiveness is limited.
On the other hand, the WaveGAN vocoder proved very useful in reducing the attack success rates in all evaluated scenarios.
While it could not fully defend the systems in a targeted attack scenario, it managed to recover most of the ground truth transcript (with 4.4\% to 15.8\% WER increase for an Espresso system under an Imperceptible attack).
Unfortunately, when the attacker can obtain a copy of the WaveGAN model to perform a joint white-box attack on WaveGAN and ASR, the defense's performance degrades but seems to be still able to prevent the attacker's target phrase recognition.
We believe our contribution will be useful as a reference for future studies of adversarial defenses for ASR systems and highlights re-synthesis (and possibly speech enhancement) as promising research directions for that purpose.


%

\appendices




\ifCLASSOPTIONcaptionsoff
  \newpage
\fi


\bibliography{spl}
\bibliographystyle{IEEEtran}

%








\end{document}